\documentclass[preprint,tightenlines,showpacs,showkeys,floatfix,nofootinbib,
superscriptaddress,fleqn]{revtex4}
\usepackage{graphicx}
\usepackage{bm}
\usepackage{dcolumn}
\usepackage{amsmath}
\usepackage{amsfonts}
\usepackage{amssymb}

\begin{document}
\preprint{INHA-NTG-03/2008}
\title{Vector and axial-vector structures of the $\Theta^+$} 
\author{Hyun-Chul Kim}
\email[E-Mail: ]{hchkim@inha.ac.kr}
\affiliation{Department of Physics, Inha University, Incheon 402-751,
Republic of Korea}
\author{Tim Ledwig}
\email[E-Mail: ]{Tim.Ledwig@tp2.ruhr-uni-bochum.de}
\affiliation{Institut f\"ur Theoretische Physik II, Ruhr-Universit\" at
  Bochum, D--44780 Bochum, Germany}
\author{Klaus Goeke}
\email[E-Mail: ]{Klaus.Goeke@tp2.ruhr-uni-bochum.de}
\affiliation{Institut f\"ur Theoretische Physik II, Ruhr-Universit\" at
  Bochum, D--44780 Bochum, Germany}
\date{May 2008}
\begin{abstract}
We present in this talk recent results of the vector and axial-vector 
transitions of the nucleon to the pentaquark baryon $\Theta^{+}$, 
based on the $SU(3)$ chiral quark-soliton model. The results are
summarized as follows: $K^{*}N\Theta$ vector and tensor coupling 
constants turn out to be $g_{K^{*}N\Theta}\simeq 0.81$ and
$f_{K^{*}N\Theta}\simeq0.84$, respectively, and the $KN\Theta$
axial-vector coupling constant to be $g_{A}^*\simeq 0.05$.  As a
result, the total decay width for $\Theta^+\to NK$ becomes very small:
$\Gamma_{\Theta\to NK}\simeq 0.71$ MeV, which is consistent with the
DIANA result $\Gamma_{\Theta\to NK}=0.36\pm0.11$ MeV.
\end{abstract}

\maketitle

\section{Introduction}
The pentaquark baryon $\Theta^+$ has been one of the most important 
issues in hadron
spectroscopy~\cite{Diakonov:1997mm,Nakano:2003qx} (See recent reviews,
for example, Refs.~\cite{Hicks:2005gp,Goeke:2004ht}).  More recently,    
the CLAS experiments reported null results of 
the
$\Theta^+$~\cite{Battaglieri:2005er,McKinnon:2006zv,Niccolai:2006td,DeVita:2006ha}, 
so that its existence has been doubted.  On the other hand, the DIANA
collaboration announced very recently the formation of a narrow
$pK^{0\text{ }}$ peak with mass of $1537\pm2$ MeV/$c^{2}$ and width of
$\Gamma=0.36\pm0.11$ MeV in the $K^{+}n\rightarrow K^{0}p$
reaction~\cite{Barmin:2006we}.  Since several new experiments for the
$\Theta^{+}$ are under way~\cite{Hotta:2005rh,Miwa:2006if,Nakano}, 
it is too early to conclude the absence of the $\Theta^+$.  However,
the total cross section for the $\Theta^+$ photoproduction should be
quite small, as already indicated from the CLAS results.  The
implications of these experimental results may be theoretically
interpreted as small coupling strengths of $K^*N\to\Theta^+$ and
$NK\to\Theta^+$ transitions.  Thus, in the present talk, we will show
that these coupling constants are actually very small, using the
chiral quark-soliton model ($\chi$QSM). 
\section{$K^*N\to\Theta^+$ and $KN\to\Theta^+$ 
  transition form factors} 
The vector and axial-vector transition form factors of the
$N\to\Theta^+$ are defined in the following matrix elements:
\begin{eqnarray}
\langle B_{\Theta^+} | \bar{\rm u}\gamma_\mu {\rm s} |
B_N\rangle &=& 
\overline{u}_{\Theta^+} \left[\gamma_\mu
F_1^{*} (Q^2) +   i\frac{\sigma_{\mu\nu}
 q^\nu F_2^{*} (Q^2) }{M_N+M_{\Theta^+}} + \frac{q_\mu F_3^*
 (Q^2)}{M_N+M_{\Theta^+}} \right] u_{N}, \cr
\langle B_{\Theta^+} | \bar{\rm u}\gamma_\mu\gamma_5 {\rm s} |
B_N\rangle &=& 
\overline{u}_{\Theta^+} \left[\gamma_\mu
g_1^{*} (Q^2) +   i\frac{
 q_\mu g_2^{*} (Q^2) }{M_N+M_{\Theta^+}} + \frac{P_\mu g_3^*
 (Q^2)}{M_N+M_{\Theta^+}} \right]\gamma_5 u_{N},
\label{general} 
\end{eqnarray}
where $Q^2=-q^2$ is the square of the four momentum transfer
$q^2=(p-p')^2$ with positive definiteness in the space-like region.
The $M_{\Theta^+}$ and $M_N$ denote the masses of the 
nucleon and $\Theta^+$, respectively.  The $\overline{u}_{\Theta^+}$
and $u_N$ are the corresponding Dirac spinors, respectively.  The
$q_\mu$ and $P_\mu$ stand, respectively, for the four momentum
transfer ($q=p'-p$) and the sum of the momenta ($P=p'+p$).  $F_i^*$
and $g_i^*$ are, respectively, vector and axial-vector form factors
that are real and functions of $q^2$.  Since the mass difference
between $M_{\Theta^+}$ and $M_N$ are not small, the $F_3^*$ and
$g_3^*$ do not vanish.  However, they do not play any important role
in determining the coupling constants for the $K^*N\Theta$ and
$KN\Theta^+$ vertices.  In fact, the third form factor $F_3^*$ in the
vector channel is related to the scalar form factor, i.e. to
the $\kappa^*$ couplings.

In the vector channel, the time and space components of the current
are directly related to the Sach-type form factors, $G_E^*$ and
$G_M^*$ as follows:
\begin{eqnarray} 
\langle   B_{\Theta^+} |\bar{\mathrm{u}}\gamma_4 \mathrm{s} |
B_N\rangle & = & G_E^* (q^2) \delta_{ss'},\cr
\langle  B_{\Theta^+}|\bar{\mathrm{u}}\gamma_k \mathrm{s} |
B_N\rangle & = & \frac{i}{2M_N} \epsilon_{klm}
(\sigma^l)_{ss'} q^m G_M^* (q^2).
\label{eq:gem}
\end{eqnarray}
We will compute these two form factors $G_E^*$ and $G_M^*$, and
axial-vector transition form factors within the framework of the SU(3)
$\chi$QSM.  In the electric transitions, we find that $G_E^*$ at
$Q^2=0$ is determined by the generalized Ademollo-Gatto (AG) theorem,
i.e., it vanishes in the chiral limit but acquires a small
contribution from the mixing angle between the octet and anti-decuplet
representations or from the wave-function corrections:
$G_E^*(0)=\sqrt{3}c_{\overline{10}}^n$.    
The detailed formalism and results can be found in Ref.~\cite{Ledwig}.   

The $K^*N\Theta^+$ vector and tensor coupling constants are determined
by the vector meson dominance (VMD).  Using the current-field identity 
\begin{equation}
  \label{eq:vdm1}
J_\mu^{4+i5} = \bar{\mathrm{u}}\gamma_\mu \mathrm{s} =
\frac{m_{K^*}^2}{f_{K^{*}}}K_\mu^*, 
\end{equation} 
where $m_{K^*}$ denotes the mass of the vector meson $K^*$.  The
$f_{K^{*}}$ stands for the generalized $K^*$ meson coupling constant 
which can be determined from the $\rho$ meson coupling constant,  
one can relate the electromagnetic transition form factors $G_E^*$ and
$G_M^*$ to the $K^*N\Theta^+$ vector and tensor coupling constants
$g_{K^*N\Theta^+}$ and $f_{K^*N\Theta^+}$.  Note that the tensor
coupling constant has been often neglected in the reaction
calculations for the $\Theta^+$
photoproduction~\cite{Oh:2003kw,Nam:2004xt}.  Ref.~\cite{Kwee:2005dz}
has kept the tensor coupling constant finite but dropped out the vector
coupling constant. 
\section{Results and discussion}
In this Section, we will briefly discuss the results of the coupling
constants.  In the following, we calculate them 
with the constituent quark mass $M=420$ MeV.  The strange current
quark mass is treated perturbatively, so that it is taken into account
to linear order ${\cal O}(m_{\mathrm{s}}^1)$.  The $m_{\mathrm{s}}$ is
selected to be $m_{\mathrm{s}}=180$ MeV which is the best value for
the mass-splitting. 

We can extract the $K^*N\Theta^+$ vector and tensor coupling constants
from the $K^*N\to\Theta^+$ electromagnetic transition form
factors with the help of the VMD:
\begin{equation}
g_{K^{*}n\Theta}=f_{K^{*}}\,\,
G_{E}^*\,\,\,\,\,\,\,\,\,\textrm{and}\,\,\,\,\,\,\,\,\,
f_{K^{*}n\Theta}=f_{K^{*}}\,\,\Big[G_{M}^*-G_{E}^*\Big].
\label{eq:gf1}
\end{equation} 
Using the values of $f_{K^*}\simeq 5.71$,  and of $G_E^*$ and $G_M^*$,
we can easily determine the $K^*N\Theta^+$ vector and tensor coupling
constants.  Let us first determine the $g_{K^{*}n\Theta}$ in the
chiral limit.  Since $G_E^*$ at $Q^2=0$ turns out to be zero due to
the AG theorem, the $g_{K^{*}n\Theta}$ becomes
also zero as already mentioned in Ref.~\cite{Azimov:2006he}.  However,
when the $m_{\mathrm{s}}$ corrections is switched on, $G_E^*$ acquires
the wave-function corrections, again, due to the AG theorem and we get
: $G_E^*\simeq0.142$.  Thus, the $K^*$ vector coupling constant becomes 
$g_{K^{*}n\Theta}\simeq0.81$.  On the other hand, the $K^*$ tensor coupling
constant is determined to be $f_{K^{*}n\Theta}\simeq 2.91$ in the chiral
limit, while with the flavor SU(3) symmetry breaking
$f_{K^{*}n\Theta}\simeq0.84$.  This large difference can be understood by
Eq.(\ref{eq:gf1}).  The finite value of $G_E^*$ due to the AG theorem
gets $f_{K^*N\Theta^+}$ decreased.  The results in the chiral limit
are consistent with those in Ref~\cite{Azimov:2006he}.  However, we
want to emphasize on the fact that due to the AG theorem the vector
coupling constant is about twice larger than the tensor one.  The
results are shown to be much smaller than those used in the reaction
calculation~\cite{Oh:2003kw,Nam:2004xt}.  Moreover, a recent
measurement for searching the $\Theta^+$ via $K^+p\to \pi^+X$ at the
KEK concludes that either the $K^*$-exchange contribution should be
excluded or $g_{K^{*}n\Theta}$ should be quite
small~\cite{Miwa:2007xk}.  Thus, the present results are very much
consistent with the KEK measurement.  The present result is also
compatible with the CLAS null results, since the total cross section
of the $\Theta^+$ photoproduction is measured and turns out very
small.   

The axial-vector constants of the nucleon have been very well
reproduced in the $\chi$QSM~\cite{silva_axialff:2005}.  Here, we
extend the calculation for the axial-vector constants, including the
baryon anti-decuplet, in particular, the $\Theta^+$.  Taking into
account the rotational $1/N_c$ corrections, linear $m_{\mathrm{s}}$
corrections, and wave-function corrections, we obtain $g_A^*$ for
the $\Theta^+\to nK^+$: $g_A^*=0.05$, which is very small.  This
smalless of the axial-vector transition constant has very important
implications, since it will predict a very small value of the
$\Theta^+\to nK^+$ decay width.  In fact, we get $\Gamma_{\Theta^+\to
  nK^+}=0.36$ MeV.  The total decay width of the $\Theta^+$ then turns
out to be $\Gamma_{\Theta^+}=0.71$ MeV.  It is a remarkable result, since
the decay width of the $\Theta^+$ is believed to be below 1 MeV.
Actually, the recent DIANA experiment measures
$\Gamma_{\Theta^+}=0.36\pm0.11$ MeV, which is very close to our
result.

\section{Summary and conclusion}
In this talk, we have presented recent results of the vector ($K^*)$
and axial-vector transition coupling constants, based on the chiral
quark-soliton model with the help of the vector meson dominance.  We
have learned that the $K^*N\to\Theta^+$ electric transition form
factor $G_E^*$ at $Q^2=0$ is soley controlled by the generalized
Ademollo-Gatto theorem.  As a result, the $K^*$ vector coupling
constant turns out to be very small.  The $K^*$ tensor coupling
constant is even about twice smaller than the vector coupling
constant.  The results are consistent with recent experimental data. 

The axial-vector transition ($\Theta^+\to nK^+$) coupling constant has
been also presented.  Its result is very small and leads to a tiny
value of the $\Theta^+$ total decay width.  It is again consistent
with the recent DIANA data.  

The detailed works on the axial-vector transition form 
factors for the $\Theta^+$ will soon appear elsewhere.
\section*{Acknowledgments}
HChK is grateful to the organizers for the Chiral07 held in Osaka,
Japan.  The work of H.Ch.K. is supported by the Korea Research
Foundation Grant funded by the Korean Government(MOEHRD)
(KRF-2006-312-C00507).  The work is also supported by the
Transregio-Sonderforschungsbereich Bonn-Bochum-Giessen, the
Verbundforschung (Hadrons and Nuclei) of the Federal Ministry for
Education and Research (BMBF) of Germany, the Graduiertenkolleg
Bochum-Dortmund, the COSY-project J\"ulich as well as the EU 
Integrated Infrastructure Initiative Hadron Physics Project 
under contract number RII3-CT-2004-506078 and the DAAD.   

\end{document}